\documentclass[a4paper,10pt]{article}

\usepackage{graphics}

\begin{document}

\begin{center}
{\bf\large Transforming from time to frequency without artefacts}\\
R. M. L. Evans\footnote{mike.evans@physics.org}\\
{\em School of Physics and Astronomy, University of Leeds, LS2 9JT, U.K.}\\
8th October, 2009
\end{center}

\begin{abstract}
I review a simple method, recently introduced to convert rheological compliance measurements into frequency-dependent moduli. New experimental data are presented, and the scientific implications of various data conversion methods discussed.
\end{abstract}

This is an article about the sanctity of experimental data.

As the readers of this Bulletin will know better than I, there are many different ways to do rheology --- many different experimental techniques by which to measure the mechanical properties of viscoelastic materials. Some techniques measure a compliance, others measure a modulus. Some measurements are most conveniently expressed as a function of time, others as a function of frequency. For instance, in a passive microrheology experiment, the time-dependent mean-square displacement of probe particles is measured as the particles perform Brownian motion within the viscoelastic medium. That increasing function of time is a measure of the fluid's time-dependent compliance \cite{Mason97,Xu,Waigh05}. There is often no alternative to such microrheological measurements, as sample volumes may be very limited, due to expense, or difficulty in producing or refining the samples \cite{Tassieri08}. The complance is also provided by a macroscopic creep experiment where a stress step is applied to a material, and its resulting strain is followed as time proceeds. Even where sufficient material exists to use a conventional rheometer, a creep experiment is often the only feasible probe of the slowest relaxation modes of high-viscosity materials such as drilling muds or soft glassy materials including pastes and slurries. The raw data generated by those experiments look quite different from the frequency-dependent storage and loss moduli ($G'(\omega)$ and $G''(\omega)$) measured in oscillatory rheometry. Nevertheless, so long as the measurements were performed in the linear regime, the two experiments in fact probe exactly the same physical properties of the material, so that the time-dependent compliance and the frequency-dependent modulus data contain the same information. 

This can be easily understood if one imagines performing oscillatory measurements at two different frequencies simulateously --- a technique known as multi-wave rheology. Applying a high-frequency oscillatory stress superimposed on top of a low-freuqency oscillatory stress causes the material to respond at both frequencies simultaneously, and the high- and low-frequency parts of its strain can be separately measured, giving its moduli at both frequencies. The experimenter could add in more frequencies, to measure more points on the spectrum at once. Every schoolboy who has watched an oscilloscope screen as sine waves from different oscillators were added together knows that, by adding up the appropriate amounts of all the odd harmonic frequencies, one can generate a square wave. A single stress step is just a square wave of infinitely low frequency, and is therefore equivalent to performing oscillatory measurements at {\em all} frequencies at once.

Since the data from creep and oscillatory rheometry contain equivalent information, it is equally acceptable to express their results by plotting a graph of compliance versus time, or of storage and loss moduli versus frequency. It is useful to express the results from both experiments in the same format, so that they can be directly compared. In practise, the frequency-dependent moduli have become the more standard rheological measure, so the need exists to convert creep compliance data into $G'(\omega)$ and $G''(\omega)$, and therein lies a problem. How do we convert the data accurately, i.e.~without introducing artefacts? 

In principle, the conversion is absolutely straightforward, since the frequency-dependent moduli are inversely proportional to the frequency-dependent compliance. We have measurements of the {\em time-}dependent compliance, so we just have to identify how much of each frequency is contained in that time-dependent signal, i.e.~Fourier-analyse it. There are plenty of standard tools available for Fourier-analysing a signal, but the compliance signal from a step-stress experiment has several features that over-tax the approximations inherent in those tools, liberally decorating the output with artefacts. We need to go back to the drawing board, and analyse closely what is involved in processing a signal, making sure not to violate basic principles of scientific method.

\begin{figure}
	\centering
		\resizebox{76mm}{!}{\includegraphics{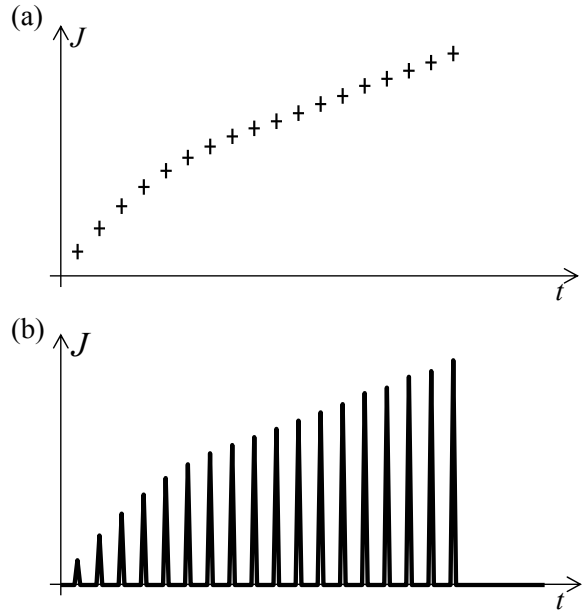}}
	\caption{\label{schematic}(a) Sketch of an example data set from experimental measurements of a time-dependent compliance $J$. (b) Na\"{\i}vely substituting the data directly into a Fourier integral (replacing the integral by a summation over the data points, $\sum_p J_p \exp(i\omega\, t_p)$) is tantamount to replacing the data by the function shown (a series of delta functions), and thus assuming that the compliance has a value of zero at all times except for those when a 
measurement was taken. That is obviously a bad assumption.}
\end{figure}

The main problem is one inherent in many experiments. We want to process (in this case, to Fourier-analyse) a signal that is a function of time. Unfortunately, we do not have access to the true function of time since, in practice, the data have three shortcomings: the measurements begin some short but non-zero time after the stress step was applied (due to the finite time resolution of the rheometer); the measurements are discrete, not continuous (also due to the temporal resolution); and the data terminate after some finite duration. The mathematical recipe for a Fourier transform requires as input a function that is defined for {\em all} values of the time $t$. So we have a dilemma: what continuous function of $t$ should we substitute into the mathematics instead of the experimental data? The knee-jerk reaction to this question is one of horrified indignation: don't substitute an artificial function; just stick the experimental data into the Fourier-transformation formula. Sadly, the Fourier-transform formula demands a continuous function, so blindly substituting-in discrete data is equivalent to replacing a data-set such as that in Fig.~\ref{schematic}a by the function in Fig.~\ref{schematic}b. Clearly, some interpolation and extrapolation is unavoidable, and we need to find a fitting function that will be truest to the data. Those who try to avoid choosing a particular function by simply substituting their data directly into the Fourier transform are unwittingly choosing to interpolate with a function such as that in Fig.~\ref{schematic}b. There could hardly be a worse choice. It would be similarly unwise to evade the choice by entering the data into some proprietary spectrum analyser. The approximation scheme would then be hidden from the scientist, but the responsibility for the resulting artefacts would remain his. Let us review the scientific validity of some other popular choices of fitting function for the compliance data.

A common approach is to fit the data to a theoretical model of viscoelastic fluids, such as the generalized Maxwell model, in which a number of Maxwell modes are added together to model relaxations in the fluid that occur at a range of different rates \cite{Bird87}. Each mode is characterized by an amplitude (how much of that mode is present) and a time scale. Thus the experimental compliance data are used to establish the values of a set of fitting parameters --- two per Maxwell mode --- and the frequency-dependent moduli of that best-fit model can then be plotted. The philosophy of such a scheme is difficult to justify, since it blurs the distinction between theory and data. Should the resulting graphs of $G'(\omega)$ and $G''(\omega)$ be regarded as the results of a model or an experiment? They can be legitimately interpreted as the output of a model, with the understanding that they therefore do not carry the weight of a true empirical observation. It would be unscientific to treat the results as measurements, since scientific methodology demands that experimental design must not allow data to be influenced by theoretical preconceptions. It must not be forgotten that experimental data, not theoretical models, are sacred.

In the same way that sine waves of different frequencies can be added together to reproduce any function (a Fourier transform), the same is true of decaying exponentials (a Laplace transform \cite{contin}); if one adds together sufficiently many decaying exponentials of appropriate amplitudes and rates, one can fit them to the experimental data with arbitrarily good accuracy. So a {\em large} series of exponentials need not be regarded as ``Maxwell modes" derived from a theoretical model, but simply as a general smooth function with sufficiently many fitting parameters to give a good approximation of the data.
Using that fitting function has the advantage that it is mathematically very easy to Fourier-transform, and thereby convert into graphs of $G'(\omega)$ and $G''(\omega)$. Philosophically, we are now of firmer ground. So long as the fitting function has enough flexibility (enough fitting parameters) to accurately reproduce the data set, it can be legitimately regarded as a true, unbiassed representation of the experimental data, so that the resulting graphs of $G'(\omega)$ and $G''(\omega)$ are scientifically valid experimental results. But how many fitting parameters are enough? How close a fit to the data is close enough?

\begin{figure}
	\centering
		\resizebox{76mm}{!}{\includegraphics{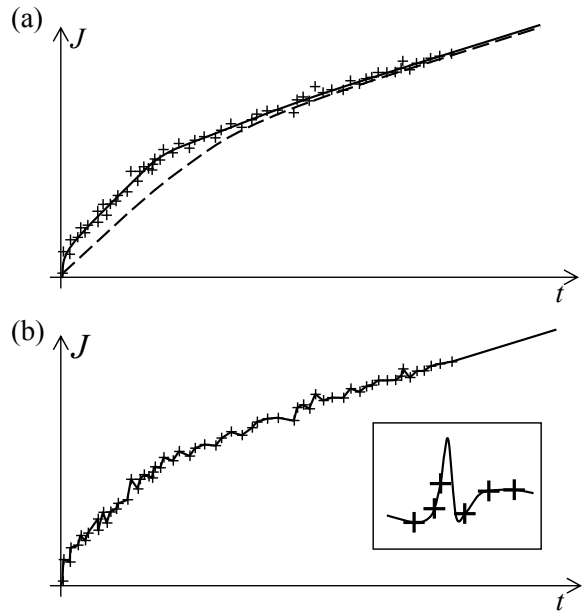}}
	\caption{\label{scatterfig}(a) Sketch of an example data set of a time-dependent compliance measurements with noise, and two alternative smooth fitting functions. (b) A picewise-linear interpolation between the data point, and also piece-wise linear extrapolations. (Inset: An alternative interpolation scheme in which the curve is unnecessarily constrained to have continuous gradient at all points.)}
\end{figure}

Intuition might suggests the following critereon. Data from the compliance experiment inevitably have some scatter, such as depicted in Fig.~\ref{scatterfig}a. If a fitting function is so inflexible (like the dashed curve) that it can't be made to pass through the middle of that scatter, then it is clearly unsuitable, as it is a biased (inaccurate) representation of the data. If, on the other hand, by fitting many parameters, one can find a smooth fitting function that passes through the middle of the noise (like the solid curve in Fig.~\ref{scatterfig}a), then it is usually assumed to be good enough. I disagree.

It is our sworn duty as scientists not to doctor data, but to report them accurately, and every graduate knows that the best way to treat experimental noise and ouliers is not to conceal them, but to present them for the reader's scruitiny. Therefore, I want to argue that the {\em only} acceptable fitting function is one that passes through {\em every} data point. Only then does the interpolating function truthfully report the results of the experiment; it is a true representation of the sacred experimental data. It is important to keep in mind the purpose of this exercise in interpolation and extrapolation, which is to allow us to Fourier-analyse the data, {\em not} to smooth them. On our quest to find a suitable interpolation scheme, we must not be beguiled by shiny things like the lovely smooth fitting function in Fig.~\ref{scatterfig}a. The assumption that the compliance should be smooth is a prejudice based on our preconceptions about fluids, but the experimental apparatus did {\em not} report a smooth function. We must put our preconceptions aside, and use the data.

So, to turn the discrete set of data points into a continuous function, suitable for Fourier analysis, we must join the dots, as in Fig.~\ref{scatterfig}b. This still leaves us with an infinite choice of functions with which to connect the data points, and one might still be tempted to construct a locally smooth function by using, for instance, a piecewise cubic or a sinc series. However, such functions have a tendency to overshoot, as depicted in the inset to Fig.~\ref{scatterfig}b. By insisting unnecessarily that the gradients match where the curves connect, in some cases one forces the curves into contrived shapes that actually lie far from the data set.
As well as passing through every data point, and thereby telling {\em the whole truth}, the chosen fitting function must tell {\em nothing but the truth}, i.e. it must have no unwarranted behaviour {\em in between} the data points, as that would introduce spurious new information that was not present in the data. The only function that reliably introduces no new features between two data points is a straight line segment. So the most truthful interpolation of the data is given by a piecewise-linear function, as sketched in Fig.~\ref{scatterfig}b. 

To turn a finite data set into a function defined for all $t$, one must not only interpolate but also extrapolate to $t=0$ and $t=\infty$, which we do with two additional straight-line segments. If the creep experiment was performed for sufficiently long to reach the terminal regime, where all elastic memory is lost, and the fluid flows with its long-time viscosity $\eta$, then the straight-line extrapolation to $t=\infty$ is uncontroversial. The final line-segment is given a gradient $1/\eta$, found by esimating the asymptote to the data by eye. Note that the human eye is the only good way to judge an asymptote to noisy or unpredictable data, as automated schemes are easily fooled. Similarly, the value $J_0$ of the compliance at $t=0$ must be carefully judged from a linear or log-linear plot. If, on the other hand, the data do not reach the terminal regime, then no amount of analysis can recover the frequency-dependent moduli from the incomplete experiment.

It turns out to be straightforward to write down the exact Fourier transform of the piecewise-linear fit, yielding a simple formula. Plugging the data points into the formula is child's play, so that some standard laborious methods for approximate conversion are rendered redundant \cite{Mason97,Plazek79,contin}. The simple formula can be found in Ref.~\cite{Evans09}, or can be downloaded for various standard data processing packages from
\texttt{www.pcf.leeds.ac.uk/research/highlight/view/4} but is also reproduced here
\begin{eqnarray*}
  G'(\omega)+i G''(\omega) 
  = & & \!\!\!\!\!\!i\omega \Big/ \bigg[ i\omega J_0
  + \left( 1-e^{-i\omega t_1}\right) \frac{\left(J_1-J_0\right)}{t_1}
  + \frac{e^{-i\omega t_N}}{\eta}	\\
  + & & \!\!\!\!\!\!\sum_{k=2}^N \left( \frac{J_k-J_{k-1}}{t_k-t_{k-1}} \right)
  \left( e^{-i\omega t_{k-1}}-e^{-i\omega t_k} \right)  \;\bigg]
\end{eqnarray*}
in terms of the two parameters $J_0$ and $\eta$, and the $N$ data points $(t_k,J_k)$. This formula will return an answer for any value of frequency $\omega$ that is substituted into it. But, of course, the data only contain information about a limited range of frequencies. Specifically, the frequency window depends on the resolution (of order $t_1$, the time of the first data point) and duration $t_N$ of the data set, so that the formula returns real, scientifically valid information for frequencies in the range $t_N^{-1}<\omega<t_1^{-1}$, and artefact-dominated rubbish for frequencies outside that range. To be clear, outside that frequency range, the function is strongly influenced by the artificial features of the fit: the straight-line extrapolation to infinity and the sharp corners of the jagged function, whereas all features (including noise) that appear on the modulus graphs {\em within} the operational frequency range are genuine features of the data.

\begin{figure}
	\centering
		\resizebox{11cm}{!}{\includegraphics{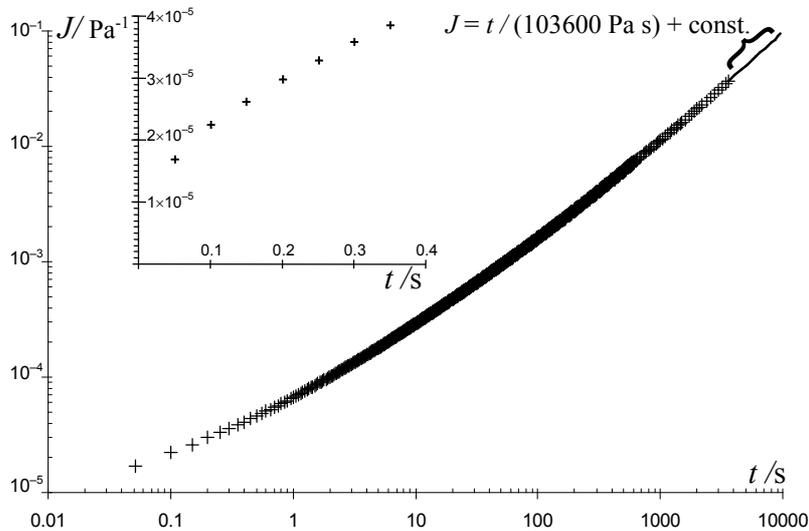}}
	\caption{\label{compliancedata} Creep compliance $J$ versus time $t$ for a polyalkylalkenylsiloxane (PAS) melt (molecular weight $M_w=5\times10^5$~g/mol, polydispersity index $M_w/M_n=1.83$) measured at room temperature using 25~mm cone-plate geometry on a stress-controlled Anton Paar MCR301 rheometer, at a stress of 200~Pa, averaged over 10 independent creep compliance runs up to time $t=600$~s. Strain rate measurements averaged from three longer runs, also at $200$~Pa, at lower temporal resolution were used to extend the range to $t=3600$~s. The linear extrapolation beyond $3600$~s is also shown, using the parameter $\eta=103600$~Pa~s. Inset: close-up on the early data points, shown on linear axes, to allow the zero-time extrapolation of $J$ to be estimated. Data courtesy of Chirag Kalelkar, Complex Fluids and Polymer Engineering Group, National Chemical Laboratory, Pune, India}
\end{figure}

\begin{figure}
	\centering
		\resizebox{10cm}{!}{\includegraphics{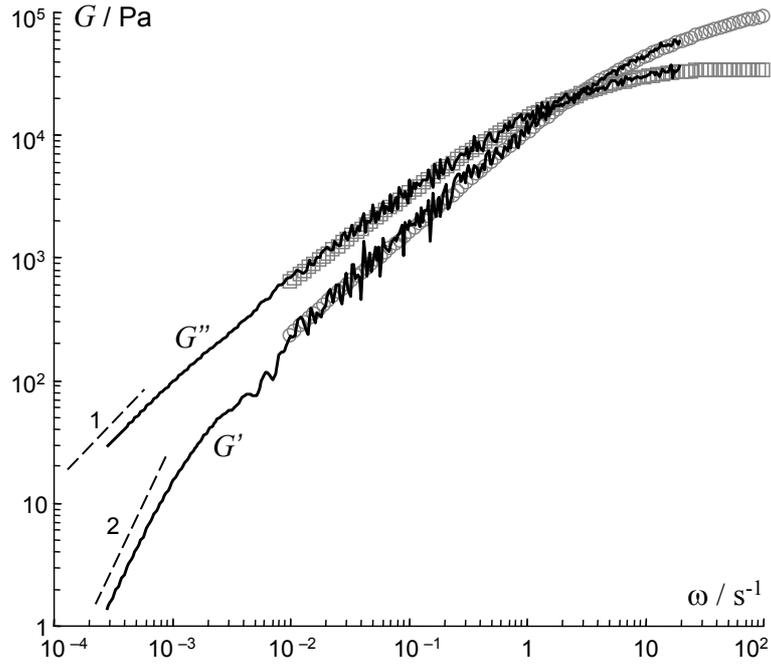}}
	\caption{\label{converteddata} Storage and loss moduli as functions of angular frequency. The continuous curves were found by plugging the compliance data of Fig.~\protect\ref{compliancedata} into the formula for $G'(\omega)$ and $G''(\omega)$. They are plotted only within the valid frequency range $t_N^{-1}<\omega<t_1^{-1}$, in this case $2.8\times10^{-4}s^{-1}<\omega<20~s^{-1}$. Also, shown for comparison are data (courtesy of Chirag Kalelkar) from oscillatory measurements (frequency sweep) on the same material ($G'$ grey circles, $G''$ grey squares).}
\end{figure}

To demonstrate the efficacy of the conversion formula, new data are presented here (courtesy of Chirag Kalelkar, Complex Fluids and Polymer Engineering Group, National Chemical Laboratory, Pune, India) for a PAS melt. Figure~\ref{compliancedata} shows the compliance measured using CP-25 geometry on an Anton Paar MCR301 rheometer, at a stress of 200~Pa, averaged over 10 independent creep compliance runs with a duration of 600~s, combined with three measurements of duration 1~hour. It is clear from the figure that the duration was sufficient to reach the terminal regime, where the complance asymptotes to a linear function (i.e.~the Newtonian regime), with viscosity $\eta=103600\pm100$~Pa~s, as demonstrated by the linear extrapolation shown in Fig.~\ref{compliancedata}. The inset to Fig.~\ref{compliancedata} shows the first few compliance data points on linear axes, allowing us to estimate the zero-time value as $J_0=(1.12\pm0.09)\times10^{-5}$~Pa$^{-1}$. Figure~\ref{converteddata} shows the results of plugging the compliance data into the formula for $G'(\omega)$ and $G''(\omega)$ (continuous curves). Also, superposed are data from oscillatory measurements (frequency sweep) on the same material. 

Note the striking agreement between the two measurements, confirming that the different experimental techniques probe the same material properties. Note also that the converted creep measurement just reaches the terminal regime (confirming that the compliance measurements were just long enough), where $G'\propto\omega^2$ and $G''\propto\omega$, at frequencies that are impractically low for oscillatory measurements. Since Fig.~\ref{converteddata} displays only the relevant frequency range $t_N^{-1}<\omega<t_1^{-1}$, this low-frequency behaviour is {\em not} an artefact of the straight-line extrapolation; it is a result dominated by the late-time data that have reached the terminal regime in the creep experiment, while the straight-line extrapolation controls features at lower frequencies than those plotted in the figure. Only if a wildy inaccurate value of the viscosity were used, that was at odds with the data, thus introducing an abrupt corner into the overall shape of the fitting function, would the resulting artefacts be so large as to leak into the operational frequency window.

The new conversion formula has already been applied to a number of different experimental systems in addition to the PAS melt reported above, including  a polyisoprene melt in cone-and-plate geometry \cite{Evans09}, magnetic microrheological measurements of F-actin solutions (a semi-felxible biopolymer) \cite{ManlioThesis}, and semidilute polyacrylamide solutions measured by both magnetic microrheology \cite{ManlioThesis} and a new protocol for obtaining microrheological creep measurements using optical tweezers \cite{ManlioUnpub}. The technique has even been applied to non-equilibrium samples of aqueous wormlike micellar solutions \cite{Willmer09}, since it does not require the sample to be at equilibrium. So long as the sample's moduli evolve only slowly --- on a much longer time-scale that the measurement --- the conversion remains valid. This makes creep the measurement of choice for slowly evolving non-equilibrium samples, as it can be performed more quickly than the equivalent oscillatory frequency sweep. Of course, the formula should not be applied to non-linear rheology or to materials that are aging rapidly on the timescale of the measurements. In both of those cases the dynamic moduli are anyway not uniquely defined, as their definitions must then take account of coupling between harmonics \cite{Wilhelm98} and of absolute time \cite{Fielding00}. As the conversion method was published only a matter of months ago \cite{Evans09}, the list of relevant literature is rather short, but growing rapidly due to the ease with which the formula can be applied.

\noindent{\small ACKNOWLEDGMENTS: Many thanks to Chirag Kalelkar for experimental data and informative discussions, and to Dave Fairhurst and Manlio Tassieri for helpful advice. RMLE is funded by the Royal Society.}



\end{document}